\documentclass[]{aa} %linenumbers
\usepackage[varg]{txfonts}
\usepackage{graphicx,color}% Include figure files
\usepackage{dcolumn}% Align table columns on decimal point
\usepackage{bm}% bold math
\usepackage{lineno, soul, ulem} 
\usepackage{hyperref,url}% add hypertext capabilities
\usepackage{xcolor}
\usepackage{array}
\usepackage{comment}

\usepackage{natbib}
\bibpunct{(}{)}{;}{a}{}{,} % to follow the A&A style

\newcommand\nustar{\textit{NuSTAR}}

\newcommand\xmm{\textit{XMM--Newton}}

\newcommand\srg{\textit{SRG}}

\newcommand{\lsim}{\mathrel{\mathop{\kern 0pt \rlap
  {\raise.2ex\hbox{$<$}}}
  \lower.9ex\hbox{\kern-.190em $\sim$}}}
\newcommand{\gsim}{\mathrel{\mathop{\kern 0pt \rlap
  {\raise.2ex\hbox{$>$}}}
  \lower.9ex\hbox{\kern-.190em $\sim$}}}

\def\erosita{eROSITA}

\def\flux{erg s$^{-1}$ cm$^{-2}$}

 \hypersetup{
    colorlinks=true,
    linkcolor=violet,
    filecolor=purple,      
    urlcolor=blue,
    citecolor=violet,
    pdftitle={ }
    }

\def\amin{\ifmmode^{\prime}\else$^{\prime}$\fi}
\def\asec{\ifmmode^{\prime\prime}\else$^{\prime\prime}$\fi}

\def\simgt{\lower.5ex\hbox{$\; \buildrel > \over \sim \;$}}
\def\simlt{\lower.5ex\hbox{$\; \buildrel < \over \sim \;$}}

\def\art{ART-XC}

\def \aap  {A\&A}

\def \apjs  {ApJS}
\def \apjl  {ApJL}

\begin{document}

\title{A wide-field X-ray search for the Geminga pulsar halo with SRG/ART-XC}

\author{Roman Krivonos\inst{\ref{inst1}}\thanks{Corresponding author. Email: \texttt{krivonos@cosmos.ru}} \and Silvia Manconi\inst{\ref{inst2},\ref{inst3}} \and  Vadim Arefiev\inst{\ref{inst1}} \and Andrei Bykov\inst{\ref{inst04} }  \and \\ Fiorenza~Donato\inst{\ref{inst5},\ref{inst6},\ref{inst7}} \and Ekaterina Filippova\inst{\ref{inst1}} \and  Alexander Lutovinov\inst{\ref{inst1}} \and Mattia Di Mauro\inst{\ref{inst5}} \and \\
Kaya Mori\inst{\ref{inst4}} \and Alexey Tkachenko\inst{\ref{inst1}} \and Jooyun Woo\inst{\ref{inst4} } }

\institute{
Space Research Institute (IKI), 84/32 Profsouznaya str., Moscow 117997, Russian Federation\label{inst1}
\and 
Laboratoire d'Annecy-le-Vieux de
Physique Théorique (LAPTh), CNRS, USMB, F-74940 Annecy, France \label{inst2} 
\and
Sorbonne Universit\'e \& Laboratoire de Physique Th\'eorique et Hautes \'Energies (LPTHE),
CNRS, 4 Place Jussieu, Paris, France
\label{inst3} 
\and
Ioffe Institute, Saint-Peterburg, Politechnicheskaya 26, 194021, Russian Federation 
\label{inst04}
\and
Department of Physics, University of Torino, via P. Giuria, 1, 10125 Torino, Italy
\label{inst6}
\and
Istituto Nazionale di Fisica Nucleare, via P. Giuria, 1, 10125 Torino, Italy
\label{inst5}
\and
European Organization for Nuclear Research (CERN), Geneva, Switzerland\label{inst7}
\and 
Columbia Astrophysics Laboratory, 550 West 120th Street, New York, NY 10027, USA\label{inst4}
}

\abstract{ 
Searches for the putative large-scale X-ray halo around the Geminga pulsar have been extensively performed using various narrow field-of-view X-ray telescopes. In this paper, we present wide-field scanning observation of Geminga with {\srg}/{\art}. Our X-ray analysis provides, for the first time, direct imaging of a $3.5^\circ\times3.5^\circ$ region in the 4$-$12~keV energy band, comparable in extent to the expected Geminga emission. The {\art} observation provides a highly uniform sky coverage without strong vignetting effects. The synchrotron X-ray halo flux was predicted using a physical model based on particle injection, diffusion, and cooling over the pulsar’s lifetime, as well as the spectral and spatial properties of the synchrotron X-ray and inverse-Compton gamma-ray emissions. The model is tuned to reproduce existing multiwavelength data from X-ray upper limits and GeV to TeV gamma-ray observations. After accounting for the high particle background and its uncertainties, no significant emission is found in the assumed source region, and X-ray flux upper limits are derived. These limits are less constraining by up to a factor of three with respect to existing results obtained with narrow field-of-view telescopes and longer exposure times. Nonetheless, we place direct and independent constraints on Geminga's ambient magnetic field strength, which are compatible with other studies.
Our methodology, including simulation for longer observation times, is applied for the first time to the wide field-of-view search for pulsar halos. {Using extensive simulations, we also show that a 68\% probability of detecting the Geminga pulsar halo can be achieved with a 20-day {\srg}/{\art} exposure for a 3~$\mu G$ magnetic field.}}

\maketitle

\bigskip

\section{Introduction}

The magnetosphere and surroundings of rapidly rotating neutron stars, i.e., pulsars, are ideal astrophysical environments for accelerating particles up to GeV--PeV energies and are observed as bright sources of nonthermal photon emission from radio to TeV energies \citep{2022ARA&A..60..495P}. 
Specifically, electron and positron pairs ($e^\pm$) are created in electromagnetic cascades in the pulsar's magnetosphere and can be further accelerated, for example, in the termination shock of the relativistic pulsar wind.
Multiwavelength signatures of these processes are accessible through the observation of pulsars' spectra, their light curves, and the extended pulsar nebula  (PWN) emission \cite{Gaensler:2006ua}. Furthermore,  after being accelerated, the $e^\pm$ are released into the interstellar medium and likely contribute to the cosmic ray density of our Galaxy,  providing a primary antimatter component of positrons to the local cosmic-ray fluxes \citep[see e.g.][]{Manconi:2020ipm,Evoli:2020szd,Orusa:2024ewq}.  
An important confirmation of this idea is the observation of a few-degree extended gamma-ray emission around one of the closest and most powerful pulsars, the Geminga pulsar \citep{Lopez-Coto:2022igd,Liu:2022hqf,Fang:2022fof,Amato:2024dss}.
When scaled to the source distance (about 250 pc), this demonstrates that $e^\pm$ populate a halo of a few tens of parsecs around the central pulsar. These particles produce gamma rays through inverse Compton emission on the interstellar radiation fields and have escaped the PWN. This phenomenon, called a ``TeV halo'' or simply a ``pulsar halo'', is supported by observations from Milagro, HAWC, HESS, LHAASO, and Fermi-LAT \citep{2009ApJ...700L.127A,HAWC:2017kbo, HESS:2023sbf, LHAASO:2021crt,DiMauro:2019yvh} for Geminga and a few other sources at TeV energies \citep{LHAASO:2021crt}. 
The $e^\pm$ populating such halos are expected to produce a complementary signature at X-ray energies, through synchrotron emission, via interaction with the ambient magnetic field. 
However, searches for the X-ray synchrotron counterpart of the GeV-TeV emitting $e^\pm$ have so far yielded null results \citep{Manconi:2024wlq,Khokhriakova:2023rqh,VERITAS:2025xjd}, resulting in strong constraints on the ambient magnetic field strength. Current X-ray observations do not cover the full angular extension of pulsar halos, specifically the observed Geminga emission at TeV energies. The typical field of view (FOV) of current X-ray observatories is limited to a few tens of arcseconds around Geminga and can be extended to a few degrees only in the hard X-ray band with NuSTAR by accounting for stray-light photons \citep{Manconi:2024wlq}. For these reasons, a wide-field search for the X-ray counterpart of the Geminga halo emission offers a complementary view and could potentially be more sensitive to the large-scale, diffuse X-ray counterpart of TeV pulsar halos. 

Recently,  X-ray emission at 0.2-2 keV, extending up to 0.2 degrees, has been reported from the putative synchrotron halo of the Monogem pulsar using {\srg}/{\erosita} performance verification data obtained in pointing mode \citep{Niu:2025sdq}; see, however, \cite{Khokhriakova:2025rhn}.
This detection would be consistent with previous null detections of the same object with XMM-Newton and {\srg}/{\erosita} surveys \citep{Liu:2019sfl,Khokhriakova:2023rqh}. 
The properties of the observed emission provide information on the strength and spatial structure of the underlying magnetic field, hinting at a nonuniform magnetic field structure. These results demonstrate that wide FOV observations are a promising avenue for elucidating particle acceleration and transport in pulsar halos. 

In this paper, we search for the putative X-ray halo of the Geminga pulsar using {\srg}/{\art} observations, covering an area of the sky comparable in size to the expected Geminga emission. These data provide a unique wide-field dataset compared with previous narrow FOV studies. Dedicated {\srg}/{\art} observations of Geminga complement the {\srg}/{\erosita} study by \cite{Khokhriakova:2023rqh}, providing harder energy coverage that is less biased by Galactic absorption and by contributions from soft diffuse emission \citep[mainly nearby Monogem Ring supernova remnant,][]{1996ApJ...463..224P}. Despite the non-detection of the Geminga pulsar halo in the current one-day {\art} observation, we demonstrate that an additional exposure can provide a potential detection or stronger constraints on Geminga's ambient magnetic field.

Our paper is organized as follows. Sect.~\ref{sec:model} describes a benchmark model for the spectral and angular properties of the Geminga halo, from GeV to TeV energies, and the synchrotron emission from the pulsar halo in the X-ray band. In Sect.~\ref{sec:obs}, we provide details of the {\srg}/{\art} scanning observations, data reduction, and spatial fitting procedure. Section~\ref{sec:results} presents the resulting upper limit on the Geminga halo detection. Section ~\ref{sec:sim} describes simulations estimating the ability of {\srg}/{\art} to improve the upper limit obtained in the current observation. Section~\ref{sec:con} summarizes the constraints on the degree-wide Geminga pulsar diffuse emission with {\srg}/{\art} and provides suggestions for future observations.

\section{Pulsar halo model}
\label{sec:model}
To model the Geminga pulsar halo, we followed the framework presented in \cite{Manconi:2024wlq}, to which the reader is referred for further details. 
Specifically, the benchmark adopted here was tuned to describe the spectral and angular properties of gamma-ray observations of the Geminga halo from GeV to TeV energies. Its main components are described below. 
 
The $e^\pm$ are produced and accelerated up to multi-TeV energies continuously at a rate that follows the evolution of the pulsar's spin-down luminosity, $\dot{E}(t) = \dot{E}_0 ( 1+ \frac{t}{\tau_0})^{-\frac{n+1}{n-1}}$.
We assumed magnetic 
dipole braking with index $n=3$, fixing the typical decay time $\tau_0=12$~kyr and the current spin-down power to $3.2 \times 10^{34}$~erg s$^{-1}$. 
The injection spectrum of the accelerated $e^\pm$ was modeled using a power law with slope $\gamma=1.85$ and an exponential cutoff $E_c=200$~TeV of the form $Q(E)=Q_0 E^{-\gamma} \exp(-E/E_c)$, where normalization $Q_0$ is related to the total rotational energy $W_0$ through a conversion efficiency $\eta=0.12$ as
\begin{equation}
\int_{0}^T dt \int_{0.1 \rm{GeV}}^{\infty} dE \, E Q(E,t) =\eta W_0\, ,
\end{equation}
in which  $T=342$~kyr is the characteristic age of the pulsar \citep[see Equations 9--12 from][]{DiMauro:2019yvh}. 

The choice of the injection spectral index $\gamma$ is consistent with the photon indexes of the X-ray synchrotron radiation measured by Chandra in the axial tail A region of Geminga \citep[see ][ for details]{2017ApJ...835...66P}, which may trace the escaping accelerated pairs. Indeed, e$^{\pm}$ pairs accelerated in the Geminga PWN could escape into the halo from this region, where photon indices of 1.4$-$1.6 measured by Chandra are produced by accelerated pairs with a power-law momenta distribution of index $\gamma= 1.8 - 2.2$.

After acceleration, $e^\pm$ begin to propagate in the pulsar surroundings and are subject to diffusion and energy losses, mainly by synchrotron and inverse Compton radiation in the ambient magnetic and photon fields, respectively.
Inverse Compton energy losses were computed using the interstellar radiation field from \cite{Vernetto:2016alq}. This process produces photons at GeV-TeV energies.
The same population of energetic $e^\pm$ pairs is expected to emit synchrotron X-rays in the keV band. This lower-energy counterpart of extended gamma-ray halos remains undetected to date, but its observation is crucial, for example, to constrain the properties of the ambient magnetic field in which $e^\pm$ propagate, as well as high-energy particle propagation models. 
For simplicity, the synchrotron emission was computed considering a uniform, random ambient magnetic field of strength $B$ extending at least to the scale of the observed gamma rays. This approach provided a means to explore values compatible with, or higher than, the upper limits derived in \cite{Manconi:2024wlq}, since the full angular extent of a possible X-ray counterpart remains unexplored.

The approximation of a uniform random field may not apply if the large-scale magnetic field around Geminga has a complicated spatial structure, for example, in the case of anisotropic diffusion \citep{Liu:2022hqf}. This could affect the properties of the predicted synchrotron emission, such as the angular distribution of the X-rays around the source, as seed e$^{\pm}$ would diffuse faster along the mean direction of the magnetic field, as quantified in \cite{Wu:2024ugx}. However, these modifications are strongly model dependent and are expected to give rise to differences in the normalization of the signal that are likely degenerate with other model parameters. In the absence of a robust synchrotron halo detection with X-rays, we considered the uniform magnetic field case for simplicity, leaving more complicated structures to further investigations. 

The morphological and spectral properties of the nonthermal emission of energetic $e^\pm$ strongly depend on how particles are injected and transported in the vicinity of the pulsar. 
Currently, three main phenomenological models have been proposed to explain the properties of observed gamma rays, namely suppressed diffusion, ballistic propagation, and anisotropic diffusion (see  \cite{Liu:2022hqf} for a review). 
We considered the first scenario, in which particle diffusion is described by a diffusion coefficient of the form $D(E)=D_0 (E/ 1 \rm{GeV})^{\delta}$, where $\delta=0.33$ and $D_0=10^{26}$~cm$^2$s$^{-1}$, suppressed by a factor of about 500 relative to the mean value inferred for the rest of the Galaxy. Apart from being the benchmark scenario used in much of the literature, suppressed diffusion can successfully explain the observed properties of the Geminga halo and other candidate halos.

As discussed in \cite{Manconi:2024wlq}, possible two-zone diffusion scenarios and the proper motion of the Geminga pulsar do not affect the synchrotron signal from Geminga in the range of a few to tens of keV. These effects are more significant for older populations of $e^\pm$ emitting at GeV via inverse Compton scattering, rather than the X-ray and TeV-emitting populations, which have had the time to propagate away from the acceleration site. 
Given our interest in the pulsar's halo diffuse emission, which is observed at TeV gamma rays to be extended a few degrees around the pulsar (corresponding to a physical size of a few tens of parsecs), the emission from the pulsar and its bow-shock nebula was not modeled here. 

\begin{figure}
    \centering
    \includegraphics[width=\linewidth]{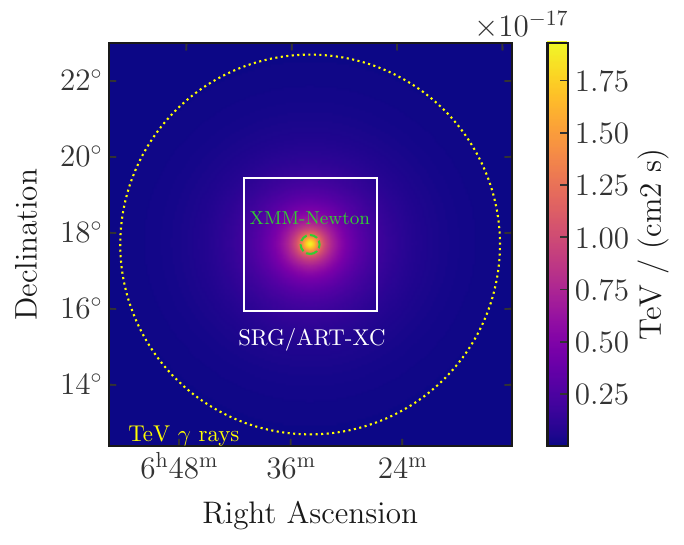}
    \caption{
    Model sky map $M_{\rm halo}$ of the Geminga X-ray halo flux coming from synchrotron emission as integrated in the 4–12 keV band, for $B=3\mu$G. The field of view of the SRG/ART-XC observation is reported with a white square, and compared to the XMM-Newton field of view from \cite{Manconi:2024wlq} (green dashed circle), as well as with the extension of the TeV gamma rays observed by HAWC (dotted yellow circle). }\label{fig:modelmap_map}
\end{figure}

Using the benchmark model outlined above, we created an intensity map by integrating within the 4$-$12 keV energy band and varying the magnetic field in the interval  $2-50~\mu$G. These physical models represent the M$_{\rm halo}$ sky maps, which are compared in the following sections with the one-day real-data observations, while in Section~\ref{sec:sim} they are used to estimate the sensitivity for a detection using longer observations. The resolution of the model sky map is $18''$. 
 Figure \ref{fig:modelmap_map} shows an example for $B=3 \mu$G. The color map represents the synchrotron flux from the Geminga pulsar halo, obtained with the benchmark model described above and integrated over the 4-12 keV energy range. The angular extension of the emission is compared with the {\xmm}, {\srg}/{\art} FOVs (green and white lines), as well as with the observed extension of the Geminga halo in gamma rays (dotted yellow line). 
By comparing the extensions in the sky of the Geminga halo synchrotron emission with the X-ray telescope FOV, the potential sensitivity of {\srg}/{\art} to a significant portion of the source's expected physical extent is evident. 

\begin{figure}
    \centering
    \includegraphics[width=\linewidth]{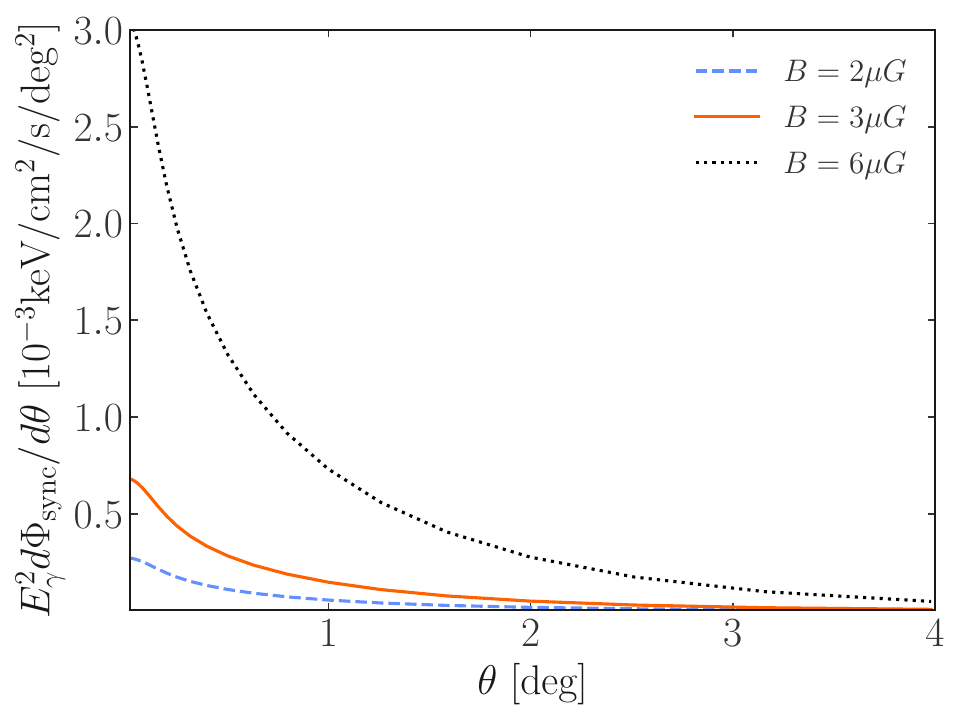}
    \caption{
    Model prediction of the surface brightness of the Geminga X-ray halo flux from synchrotron emission, integrated over the 4$-$12~keV band within four degrees of the center. The magnetic field strength varies from $2~\mu$G (dashed blue line) to $6~\mu$G (dotted black line).}\label{fig:modelmap_sb} 
\end{figure}

A complementary view of the expected emission properties within the SRG/ART-XC observation window is presented in Figure \ref{fig:modelmap_sb}, which illustrates the surface brightness, i.e., the synchrotron flux from the Geminga pulsar halo as a function of the distance from the center. 
Depending on the magnetic field value $B$, which is varied here from $2~\mu$G to $6~\mu$G, a decrease of a factor of approximately ten is predicted within the inner two degrees. 

\section{\art\ observations of Geminga}
\label{sec:obs}

In this section, we describe the observations of the Geminga region performed with {\art}. We first detail the data collection strategy and then describe the data reduction methodology and the image fitting procedure.

\subsection{Data collection}
The region of Geminga was observed on 16 April 2023 with the {\it Mikhail Pavlinsky\/} Astronomical Roentgen Telescope -- X-ray Concentrator (\art, \citealt{2021A&A...650A..42P}), one of two X-ray telescopes on board the \srg\ observatory \citep{srg}, launched on 13 July 2019, from the Baikonur Cosmodrome.

The \art\ instrument is an X-ray telescope assembly containing seven independent X-ray telescope modules. It covers the 4$-$30~keV energy range and extends the energy response of the other instrument on board \srg\ observatory, \erosita\ \citep{2021A&A...647A...1P}, which is sensitive in the 0.2$-$8~keV band.

The Geminga region was observed with a total exposure of  82~ksec.\footnote{The \erosita\ telescope was not operating during this period.} in scanning mode, specially developed for X-ray imaging of extended objects with angular sizes greater than the FOV of the {\srg} instruments \citep{srg,2022MNRAS.510.3113K,2024MNRAS.529..941S}. A single scan with a size of $3.5\times3.5$ degrees, centered on the Geminga pulsar position, was performed with a scan step of $4^{\prime}$. This step is optimized for the vignetting function of the \art\ telescope to ensure a uniform exposure. As a result, the vignetting-corrected exposure reached ${\sim}1800$ seconds at each point.

\subsection{Data analysis}
\label{sec:data}

We reduce the raw \art\ telemetry data using the data analysis tools developed in IKI\footnote{Space Research Institute of the Russian Academy of Sciences, Moscow, Russia}, as outlined in \citet{2021A&A...650A..42P}. Using the \textsc{ artpipeline} software, we produced clean, calibrated event lists for each of the telescope modules and spacecraft attitude data. The cleaned science data were then processed with \textsc{artproducts} to obtain sky images, exposure, and particle background maps.

\begin{figure}
    \centering
    \includegraphics[width=\linewidth]{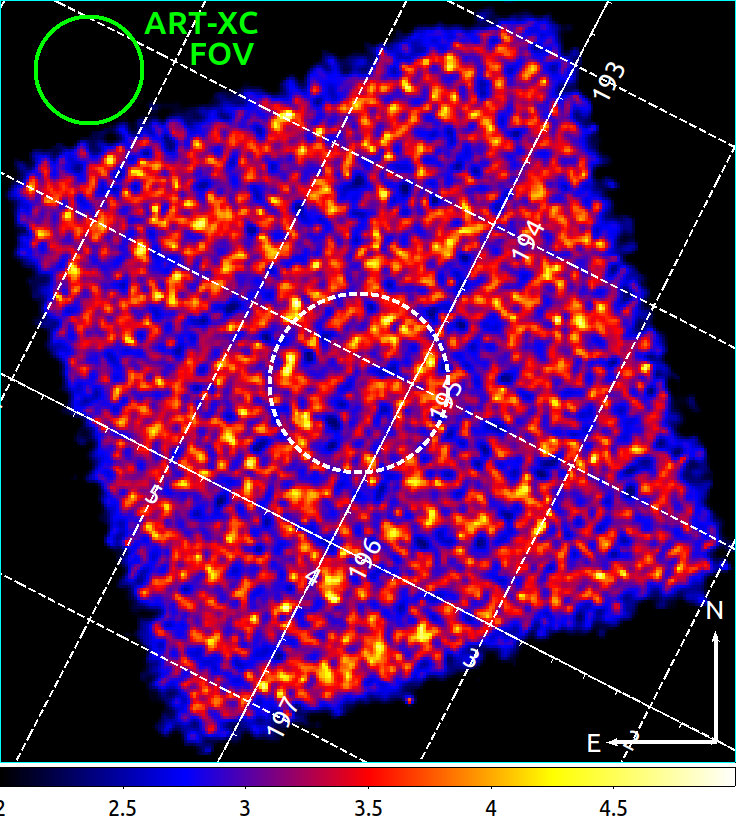}
    \caption{Sky image of the Geminga pulsar region in the 4$-$12~keV obtained with {\art}, shown in units of counts~pix$^{-1}$.  The image is smoothed with the \texttt{dmimgadapt} task from CIAO-4.15 using a Gaussian kernel. The corresponding exposure map is highly uniform, with ${\sim}1800$~seconds per pixel. The position of the Geminga pulsar is marked by a white circle of $1^{\circ}$ diameter. The grid is displayed in Galactic coordinates. The green circle denotes the \art\ FOV, $36'$ in diameter.}
    \label{fig:map}
\end{figure}

We combined the \art\ cleaned event lists for all seven telescope modules into a single sky mosaic in the 4$-$12~keV  energy band, which is the most sensitive energy range for detecting point-like and extended X-ray sources, given the energy dependence of the \art\ effective area \citep{2016SPIE.9905E..1JP}.

Figure~\ref{fig:map} shows an adaptively smoothed sky image of Geminga obtained with the one-day \art\ data in the 4$-$12~keV energy range. The data are presented in Galactic coordinates and in units of counts~pix$^{-1}$. 
The position of the Geminga pulsar is indicated by a white circle with a diameter of $1^{\circ}$. The point-like emission of the Geminga pulsar is not detected.
The  corresponding upper limit of $3\times10^{-12}$\flux\ ($2\sigma$) in the 4$-$12~keV band is consistent with a 4$-$12~keV flux of $4\times10^{-13}$\flux\ derived from a joint {\nustar} and {\xmm} power-law fit above 3 keV, as reported by \cite{Mori:2014gaa}.

In principle, the data contain uniformly distributed photons from the particle background and the cosmic X-ray background (CXB). 
In addition, a putative extended emission of Geminga may be present, which is expected to follow the physical halo model described in Section~\ref{sec:model} (M$_{\rm halo}$, Fig.~\ref{fig:modelmap_map}) and to be centered on the midpoint of the adopted scan.
This pilot \art\ observation provides a clean data set with uniform sky coverage, which we used to search for the Geminga halo and to guide the feasibility of future observations.

\subsection{Image fitting procedure}
\label{sec:sherpa}

We searched for X-ray emission from the Geminga pulsar halo using wide-field observations with {\art}. Since the instrument operates in photon counting mode, we first converted the sky intensity map of the halo model to the {\art} pixel frame. Next, we converted the model intensity (erg~s$^{-1}$~cm$^{-2}$~pix$^{-1}$) to counts (cts~s$^{-1}$~cm$^{-2}$~pix$^{-1}$), assuming an absorbed power-law model with photon index $\Gamma=2$. We verified that this photon index adequately approximates the trend of the physical models over the considered energy range. The absorbing hydrogen column density of $3.57\times 10^{21}$~cm$^{-2}$ was adopted from H14PI project \citep{2016A&A...594A.116H}. 

To extract information on the putative contribution from the Geminga X-ray halo, we constructed a model for the number of counts within the observed region of interest (ROI). 
This model includes a uniform background, which comprises a combination of the particle background and CXB, and the Geminga halo model, $M_{\rm halo}$.
We followed a widely used statistical paradigm, which considers the expected number of counts in each pixel, ${<}N_{\rm pix}{>}$, predicted by the model, and observed counts, $N_{\rm pix}$, in the data map. 
The model incorporates two free parameters (N$_{\rm bkg}$, S$_{\rm halo}$), which are optimized on the data map. Specifically,
we estimated the best-ft model parameters by maximizing the likelihood that the observed count map is described by the halo model. To do this, we used the \textsc{SHERPA} modeling and fitting package \citep{2001SPIE.4477...76F,2024ApJS..274...43S}, a part of the \textsc{CIAO} software \citep{ciao}, to construct a 2D model and fit it to the data. 

As described above, our setup included a uniform background (comprising the particle background and CXB) and a space-dependent Geminga halo model, $M_{\rm halo}$, which was convolved with the {\art} point spread function (PSF)\footnote{Given the large angular size of the Geminga pulsar halo, the knowledge of the {\art} mirror's response at large radii is crucial. We used the updated in-flight PSF calibrations made by \cite{2025ExA....59...37K}, who used the observations of the bright X-ray source Sco X-1 to calibrate the {\art} PSF shape up to ${\sim}1.5^{\circ}$.}. 
The  model setup within \textsc{SHERPA} is expressed as the sum of two contributions: \texttt{psf(M$_{\rm halo}$)*const2d.S$_{\rm halo}$*emap} + \texttt{const2d.N$_{\rm bkg}$*emap}. 
In this expression, \texttt{psf(M$_{\rm halo}$)} represents a convolution of a 2D sky map, $M_{\rm halo}$, with the PSF, which is normalized by the $S_{\rm halo}$ parameter and multiplied by the exposure map, \texttt{emap}. The background is added as \texttt{const2d.N$_{\rm bkg}$*emap}.  Technically, the fit parameters enter as \texttt{const2d} variables within the SHERPA fit.
The background is not convolved with the PSF. 

The two free parameters, $S_{\rm halo}$ and \texttt{N$_{\rm bkg}$}, are expressed in units of cts~s$^{-1}$~pix$^{-1}$.
The fit was repeated for the synchrotron model, M$_{\rm halo}$, computed for different values of the magnetic field, $B$.

For each tested $B$, the fit results for the $S_{\rm halo}$ parameter can be interpreted as follows:
\begin{enumerate}
    \item $S_{\rm halo}\approx 1$ with a small error indicates a detection of the Geminga halo emission 
    %predicted by the current model $M_{\rm halo}$;
    predicted by $M_{\rm halo}$ for that fixed $B$. 
    \item $S_{\rm halo}>1$ implies that the emission model must be amplified by a factor of $S_{\rm halo}$.
    \item 
    The non-detection of the Geminga halo is indicated by either
    a small value, $S_{\rm halo}\approx0$, not consistent with unity, or by
    %$S_{\rm halo}\approx0$ and 
    $S_{\rm halo}$ being consistent with both zero and unity within the errors. 
    In the first case, the model predicts too many counts, while in the second case, we do not detect the halo emission and cannot constrain the model $M_{\rm halo}$.
    %\item The upper limit on the Geminga halo is governed by $S_{\rm halo}\approx0$ and not consistent with unity. This case means that we do not detect halo emission and {\it can} constrain the model $M_{\rm halo}$
\end{enumerate}
The upper limit on the Geminga halo is obtained when $S_{\rm halo}\approx0$ and is not consistent with unity. This case means that we do not detect the halo emission and that we can constrain the model $M_{\rm halo}$,  and thus $B$. Based on the current pilot {\art} data set, we realistically expected a non-detection, and the ability to place upper limits.

\section{Results}
\label{sec:results}
Table~\ref{tab:bestfit} presents the best-fitting parameters of the background normalization, $N_{\rm bkg}$, and scaling factor, $S_{\rm halo}$, for the Geminga halo model, calculated for different test magnetic field strengths: 3, 10, 15, 30, and 50~$\mu$G. As seen from the table, the constant background is determined accurately;
the values found for the different magnetic fields tested are consistent and within percent-level uncertainties.
From the first line of the table, we see that the 3~$\mu G$ Geminga halo is not detected, as expected given the current statistics of the {\art} data set. The halo model with higher fields is also not detected. However, the fitting procedure can provide 1-$\sigma$ upper limits.

\begin{table}
    \centering
        \caption{Best-fitting parameters of the background normalization, $N_{\rm bkg}$, and Geminga halo model scaling factor, $S_{\rm halo}$, for different magnetic field strengths. }
    \begin{tabular}{ccc}
        M field & $N_{\rm bkg}$ & $S_{\rm halo}$ \\
        $\mu G$ & $10^{-4}$ cts s$^{-1}$ pix$^{-1}$ & \\
        \hline
         3 & $5.8669 \pm 0.0426$ & $0.91 \pm 4.07$ \\
         10 & $5.8653 \pm 0.0496$ & $ 0.074\pm 0.335$ \\
         15 & $5.8648 \pm 0.0522$ & $0.033 \pm  0.152$\\
         30 & $5.8637 \pm 0.0567$ & $ 0.009\pm 0.041$ \\
         50 & $5.8628 \pm 0.0602$ & $0.004 \pm 0.016$ \\
    \end{tabular}
    \tablefoot{Uncertainties are reported for the 68\% confidence level.
}
    \label{tab:bestfit}
\end{table}

\begin{figure}
    \centering
\includegraphics[width=\columnwidth]{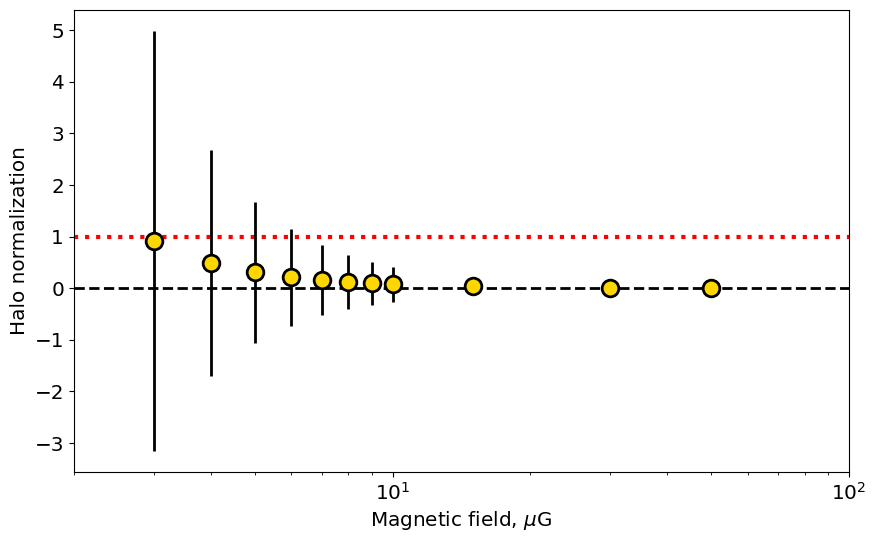}
    \caption{Geminga halo emission model scaling factor, $S_{\rm halo}$, obtained from one-day {\art} data, shown as a function of magnetic field strength.}
    \label{fig:mfield}
\end{figure}

To identify the 1-$\sigma$ upper limit on $B$, 
we calculated the model for additional field strengths below 10~$\mu G$.
Figure ~\ref{fig:mfield} shows $S_{\rm halo}$ as a function of the magnetic field strength.  According to the methodology in Section~\ref{sec:obs}, the highest magnetic field consistent with the {\art} dataset is 6~$\mu G$ at the 68\% confidence level. As shown in Fig.~\ref{fig:mfield}, the point for the Geminga halo model corresponding to 6~$\mu G$ is consistent with zero and not with unity within errors. In other words, the limits derived from the one-day {\art} scanning observation are not consistent with a magnetic field ${>}6~\mu$G.

The X-ray flux upper limits derived from the one-day {\art} observations are less constraining by a factor of three compared to existing limits obtained with narrow FOV telescopes \cite{Manconi:2024wlq}. We find it remarkable that, within just one-day observation, the physical parameters of the Geminga halo model can be constrained to values that are usually predicted by current state-of-the-art models of the Galactic magnetic field \cite{Unger:2023lob}.

\section{Simulations}
\label{sec:sim}
To estimate the ability of {\art} to improve the current upper limit on the detection of the Geminga halo emission, we performed extensive simulations covering longer observation times.

\begin{figure}[ht]
    \centering
    \includegraphics[width=\linewidth]{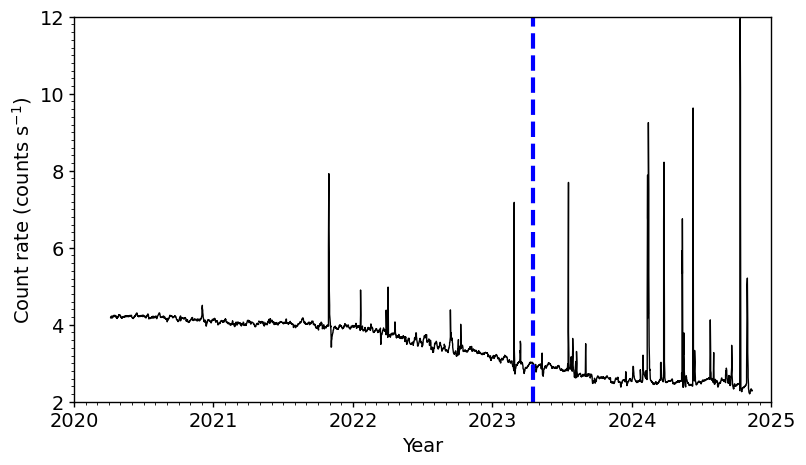}
    \caption{Background count rate measured by {\art}in the 60$-$120~keV band as a function of time in years. The vertical dashed line indicates the date of the {\srg}/{\art} scanning observation of Geminga on 16 April 2023. The figure is based on data obtained from the {\srg}/{\art} environment monitor, available at \url{https://monitor.srg.cosmos.ru/}.}
    \label{fig:l2}
\end{figure}

\begin{figure}
    \centering
    \includegraphics[width=\linewidth]{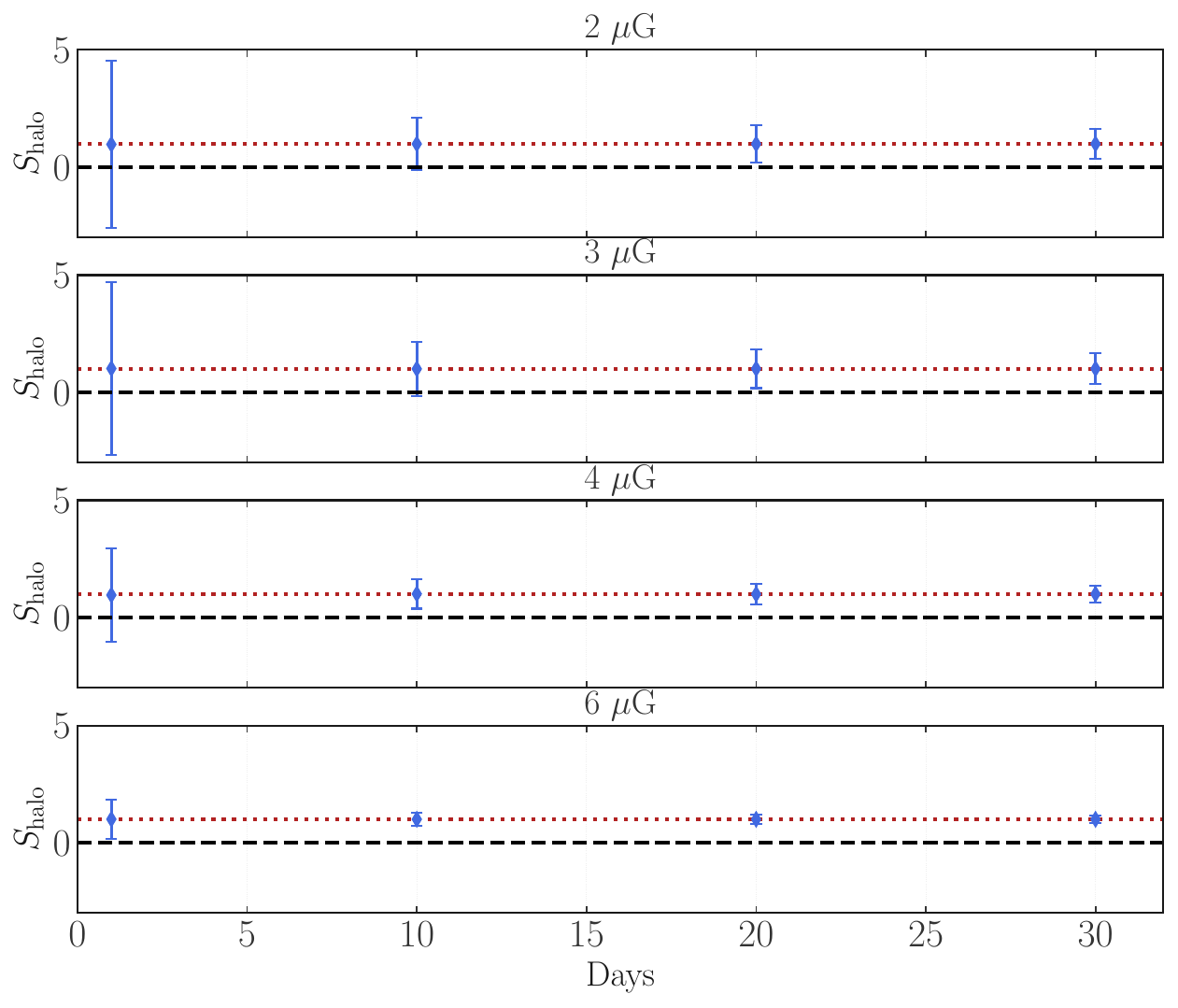}
    \caption{Simulations of Geminga halo detection at different {\srg}/{\art} exposures. The magnetic fields considered are 2, 3, 4, and 6~$\mu$G. The plot shows the Geminga emission model scaling factor, $S_{\rm halo}$, and its 1-$\sigma$ uncertainty. The dotted red line indicates $S_{\rm halo}=1$.}
    \label{fig:sim:mfield}
\end{figure}

First, we simulated the 4$-$12~keV background image based on the expected background count rate. The {\art} background at the L2 point slowly decreases with time, as a result of the current rise in solar activity.  Fig.~\ref{fig:l2} shows the evolution of the {\art} internal detector background in the 40$-$60~keV energy band, in which {\art} does not have an energy response to incoming X-ray photons. However, since the one-day observation of Geminga in 2023, the background has decreased by ${\sim}20\%$, along with increased solar flaring activity. Nevertheless, for the simulation of the 4$-$12 keV background map in 2025, we used 20\% less than the 2023 background count rate, assuming a low probability of catching a flare during the observation. We then simulated the large-scale Geminga halo emission based on a given model setup. We combined the maps and ran the 2D fitting procedure described in Sect.~\ref{sec:sherpa}, to estimate $S_{\rm halo}$. By repeating the entire procedure $10^4$ times, we obtained the distribution of $S_{\rm halo}$ values.

Figure~\ref{fig:sim:mfield} shows the mean and 1-$\sigma$ width derived from the distribution of $S_{\rm halo}$ values as a function of exposure time in days. We ran simulations for the Geminga halo model calculated for magnetic field strengths of 2, 3, 4, and 6~$\mu G$.  The first value is relevant to our recent magnetic field constraint based on the {\xmm} data \citep{Manconi:2024wlq}.

Simulation of all the explored values of the magnetic field confirms the results of the current work for the one-day Geminga observation in 2023, showing no detection of the halo and an upper limit for one-day observation at around $6\mu$G. As seen in the figure, the ten-day exposure should improve the one-day upper limit by a factor of two. Increasing the exposure time to 20 days would result in the detection of the Geminga halo emission corresponding to the model prediction for $B=3\mu$G at the 68\% probability level. Finally, we note that the Geminga observation mode can be changed from scan to grid tiling with smaller sky coverage, e.g.,$1.5^\circ\times1.5^\circ$, which is better optimized for a 2$-$3~$\mu G$ magnetic field (Fig.~\ref{fig:modelmap_sb}) and requires less exposure time than scan mode.

\section{Summary and conclusion}
\label{sec:con}

In this work, we performed a wide-field search for the putative large-scale X-ray halo around the Geminga pulsar using a scanning, one-day observation with the {\art} telescope onboard the {\srg} observatory. 
This probe observation covers a large area of the sky, comparable with the size of expected Geminga X-ray emission. Thus, compared to other X-ray searches for the Geminga halo, the {\art} provides a unique data set characterized by direct imaging in the 4$-$12~keV energy band and a highly uniform sky coverage. We developed a physical model scaled on the existing gamma-ray measurements of the Geminga pulsar halo at high energies (GeV/TeV), which predicts spatially distributed synchrotron X-ray halo flux.
This model predicts the X-ray flux in the 4$-$12~keV energy band as a function of magnetic field strength. We performed an image-fitting procedure in which the normalization of the intensity of the Geminga synchrotron halo emission and the uniform background level enter the fit as free parameters. 
We find that the model with a magnetic field of $6~\mu G$ is compatible with the {\art} data at $68\%$ confidence and therefore provides an upper limit for the magnetic field strength. The moderate constraint on the magnetic field, higher by about a factor of two compared to other state-of-the-art studies, is mainly governed by the high instrumental background.
Remarkably, within just a one-day {\art} observation, the magnetic field around the Geminga halo is constrained to values of the same order of magnitude as the expected Galactic magnetic field strength.
Additionally, using extensive simulations, we show that longer {\art} observations of Geminga can significantly improve these constraints.
A ten-day exposure can improve the upper limit on the magnetic field by a factor of two relative to the one-day observation. Observations of about 20 days or more have the potential to uncover the synchrotron emission from the Geminga pulsar halo. 

The techniques developed in this work for {\art} can be generalized to test other theoretical models
of the Geminga pulsar halo, 
in addition to the benchmark suppressed diffusion model investigated here. In such cases, it would be necessary to build a consistent spectral and spatial template. 
Furthermore, the techniques developed and applied to the {\art} data in this study are applicable to data from other wide field-of-view X-ray instruments, such as the Advanced X-ray Imaging Satellite (AXIS; \cite{Reynolds:2023vvf}).
Detecting an X-ray counterpart of the Geminga gamma-ray halo would provide a unique probe of particle acceleration and electron-positron transport around pulsars. Such a detection would inform current theoretical models, for example, regarding the structure of the magnetic field through which these relativistic particles propagate, and improve our understanding of the halo phenomenon around Galactic pulsars.

\section*{Data availability}
Figure \ref{fig:map} is available in electronic form at the CDS via anonymous ftp to cdsarc.u-strasbg.fr (130.79.128.5) or via \url{http://cdsweb.u-strasbg.fr/cgi-bin/qcat?J/A+A/}.
At the time of writing, the {\srg}/{\art} data and the corresponding data analysis software have a private status. Public access to the {\art} scientific archive will be provided in the future.

\begin{acknowledgements} 
The {\it Mikhail Pavlinsky} \art\ telescope is the hard X-ray instrument on board the \srg\ observatory, a flagship astrophysical project of the Russian Federal Space Program realized by the Russian Space Agency in the interests of the Russian Academy of Sciences. The \art\ team thanks the Russian Space Agency, Russian Academy of Sciences, and State Corporation Rosatom for the support of the \srg\ project and \art\ telescope. SM acknowledges the European Union's Horizon Europe research and innovation program for support under the Marie Sklodowska-Curie Action HE MSCA PF–2021,  grant agreement No.10106280, project \textit{VerSi} and the support of the French Agence Nationale de la Recherche (ANR), under the grant ANR-24-CPJ1-0121-01.
F.D.~and M.D.M.~acknowledge support from the Research grant {\sl TAsP (Theoretical Astroparticle Physics)} funded by Istituto Nazionale di Fisica Nucleare (INFN). M.D.M.~acknowledges support from the Italian Ministry of University and Research (MUR), PRIN 2022 ``EXSKALIBUR – Euclid-Cross-SKA: Likelihood Inference Building for Universe’s Research'', Grant No. 20222BBYB9, CUP I53D23000610 0006, and from the European Union -- Next Generation EU.
\end{acknowledgements}

\bibliographystyle{aa} % style aa.bst
\bibliography{biblio}

\end{document}